\documentclass[sigconf]{acmart}

\usepackage{subfigure}
\usepackage{url}
\usepackage{multirow}
\usepackage{algorithm}
\usepackage{algorithmic}
\usepackage{balance}

\begin{document}

\title{Contrastive Learning for Conversion Rate Prediction}
\author{Wentao Ouyang}
\affiliation{%
  \institution{Alibaba Group}
   \country{}
}
\email{maiwei.oywt@alibaba-inc.com}

\author{Rui Dong}
\affiliation{%
  \institution{Alibaba Group}
  \country{}
}
\email{kailu.dr@alibaba-inc.com}

\author{Xiuwu Zhang}
\affiliation{%
  \institution{Alibaba Group}
  \country{}
}
\email{xiuwu.zxw@alibaba-inc.com}

\author{Chaofeng Guo}
\affiliation{%
  \institution{Alibaba Group}
  \country{}
}
\email{chaofeng.gcf@alibaba-inc.com}

\author{Jinmei Luo}
\affiliation{%
  \institution{Alibaba Group}
  \country{}
}
\email{cathy.jm@alibaba-inc.com}

\author{Xiangzheng Liu}
\affiliation{%
  \institution{Alibaba Group}
  \country{}
}
\email{xiangzheng.lxz@alibaba-inc.com}

\author{Yanlong Du}
\affiliation{%
  \institution{Alibaba Group}
  \country{}
}
\email{yanlong.dyl@alibaba-inc.com}

\renewcommand{\shortauthors}{Wentao Ouyang, et al.}

\begin{abstract}
Conversion rate (CVR) prediction plays an important role in advertising systems. Recently, supervised deep neural network-based models have shown promising performance in CVR prediction. However, they are data hungry and require an enormous amount of training data. In online advertising systems, although there are millions to billions of ads, users tend to click only a small set of them and to convert on an even smaller set. This data sparsity issue restricts the power of these deep models. In this paper, we propose the Contrastive Learning for CVR prediction (CL4CVR) framework. It associates the supervised CVR prediction task with a contrastive learning task, which can learn better data representations exploiting abundant unlabeled data and improve the CVR prediction performance. To tailor the contrastive learning task to the CVR prediction problem, we propose embedding masking (EM), rather than feature masking, to create two views of augmented samples. We also propose a false negative elimination (FNE) component to eliminate samples with the same feature as the anchor sample, to account for the natural property in user behavior data. We further propose a supervised positive inclusion (SPI) component to include additional positive samples for each anchor sample, in order to make full use of sparse but precious user conversion events. Experimental results on two real-world conversion datasets demonstrate the superior performance of CL4CVR. The source code is available at https://github.com/DongRuiHust/CL4CVR.
\end{abstract}



\copyrightyear{2023}
\acmYear{2023}
\setcopyright{acmlicensed}\acmConference[SIGIR '23]{Proceedings of the 46th International ACM SIGIR Conference on Research and Development in Information Retrieval}{July 23--27, 2023}{Taipei, Taiwan}
\acmBooktitle{Proceedings of the 46th International ACM SIGIR Conference on Research and Development in Information Retrieval (SIGIR '23), July 23--27, 2023, Taipei, Taiwan}
\acmPrice{15.00}
\acmDOI{10.1145/3539618.3591968}
\acmISBN{978-1-4503-9408-6/23/07}

\settopmatter{printacmref=true}
%\fancyhead{}

\maketitle

\section{Introduction}
Conversion rate (CVR) prediction \cite{lee2012estimating,chapelle2014modeling,lu2017practical,guo2021enhanced} is an essential task in online advertising systems.
The predicted CVR impacts both the ad ranking strategy and the ad charging model \cite{zhu2017optimized,ma2018entire,pan2019predicting,pan2022metacvr}.

Recently, deep neural network-based models have achieved promising performance in CVR prediction \cite{ma2018entire,wen2020entire,guo2021enhanced,xu2022ukd}. However, deep models are data hungry and require an enormous amount of training data. In online advertising systems, although there may be millions to billions of ads, users tend to click only a small set of them and to convert on an even smaller set. This data sparsity issue restricts the prediction power of these deep models.

Contrastive learning (CL) \cite{chen2020simple,yu2022self} offers a new way to conquer the data sparsity issue via unlabeled data.
The idea is to impose different transformations on the original data and to obtain two augmented views for each sample.
It then pulls views of the same sample close in the latent space and pushes views of different samples apart in order to learn discriminative and generalizable representations.

In this paper, we propose the Contrastive Learning for CVR prediction (CL4CVR) framework, which associates the supervised CVR prediction task with a CL task. The CL task can learn better data representations and improve the CVR prediction performance.
The way to create different data augmentations highly impacts the performance of CL. In recommender systems, most data augmentation methods are ID-based sequence or graph approaches \cite{xie2022contrastive,wu2021self,yu2022self}, which do not apply to CVR prediction which is a feature-rich problem. The most relevant work is \cite{yao2021self}, which proposes feature masking for item recommendation.
The aim is that two differently masked views, each containing part of item features, can still well represent the same item.
However, feature masking does not work well for CVR prediction, because the input features are diverse, which relate to the user, the item, the context and the interaction, rather than only the item. We cannot make a good CVR prediction if we only know the target user but not the target item (which is masked).

To tailor the CL task to the CVR prediction problem, we propose embedding masking (EM), rather than feature masking, to generate two views of augmented samples. In this way, each augmented view contains all the features, except that some embedding dimensions are masked.
The CL loss will force the learned embeddings to be more representative.
We also propose a false negative elimination (FNE) component to account for the natural property in user behavior data. We further propose a supervised positive inclusion (SPI) component to make full use of sparse but precious user conversion events. Experimental results show that the proposed EM, FNE and SPI strategies all improve the CVR prediction performance.

In summary, the main contributions of this paper are
\begin{itemize}
\item We propose the CL4CVR framework, which leverages a contrastive learning task to learn better data representations and to improve the CVR prediction performance.
\item We propose embedding masking for data augmentation that is tailored to feature-rich CVR prediction.
\item We propose a false negative elimination component and a supervised positive inclusion component to further improve the contrastive learning performance.
\end{itemize}
\vskip -2pt

\begin{figure}[!t]
\centering
\includegraphics[width=0.36\textwidth, trim = 0 0 0 0, clip]{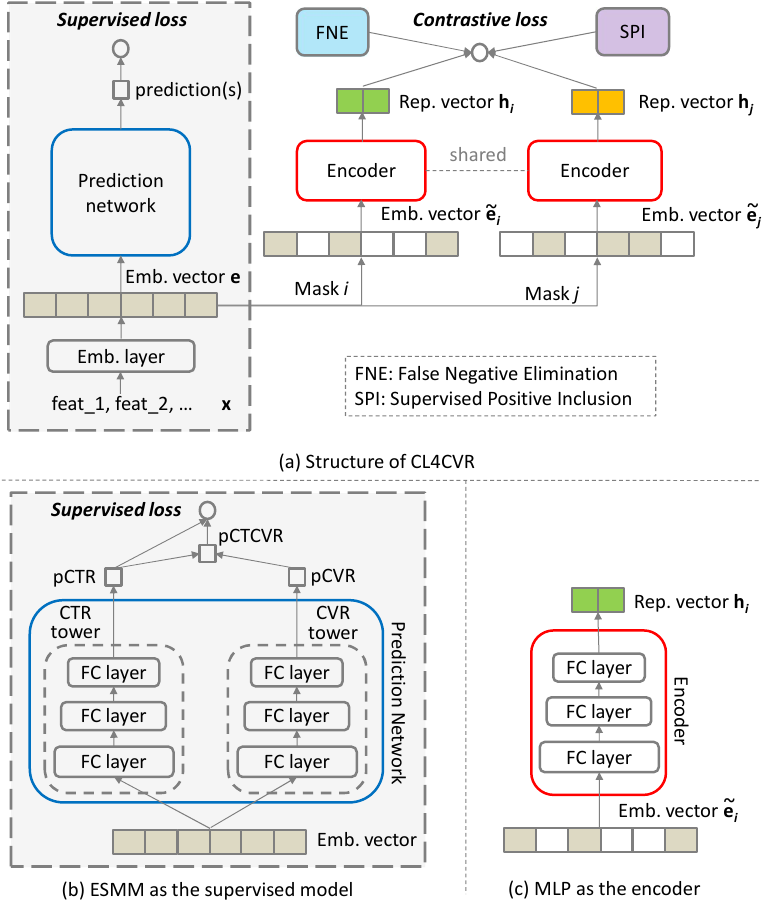}
\vskip -8pt
\caption{(a) Structure of the CL4CVR framework. (b) ESMM as the supervised prediction model. (c) MLP as the encoder.}
\vskip -10pt
\label{fig_model}
\end{figure}

\section{Model Design}
We propose the CL4CVR framework that combines contrastive learning (CL) with supervised learning (SL) to improve the performance of CVR prediction. The structure is shown in Fig. \ref{fig_model}(a).

\subsection{Problem Formulation}
In typical advertising systems, user actions follow an impression $\rightarrow$ click $\rightarrow$ conversion path.
Denote the input feature vector as $\mathbf{x}$, which contains multiple fields such as user ID, gender, age group, ad ID, ad title, ad industry, city, OS, etc. If a click event occurs, the click label is $y=1$, otherwise, $y=0$. If a conversion event occurs, the conversion label is $z=1$, otherwise, $z=0$.
The (post-click) CVR prediction problem is to estimate the probability
$\hat{z} = p(z=1|y=1, \mathbf{x})$.

\subsection{Supervised Prediction Model}
Our focus in this paper is on the design of the CL task, and we use existing CVR prediction model as the SL task.
In particular, we use ESMM \cite{ma2018entire} as the supervised prediction model because of its popularity and versatility.
More sophisticated models such as ESM$^2$ \cite{wen2020entire}, GMCM \cite{bao2020gmcm} and HM$^3$ \cite{wen2021hierarchically} require additional post-click behaviors (e.g., favorite, add to cart and read reviews), which are not always available in different advertising systems.

Fig. \ref{fig_model}(b) shows the structure of ESMM. It has a shared embedding layer, a CTR tower and a CVR tower. Assume there are $N$ samples in a mini-batch. Denote the predicted CTR as $\hat{y}$ and the predicted CVR as $\hat{z}$, the supervised loss is defined as
\begin{equation}
L_{pred} = \frac{1}{N} \sum_{n=1}^N l(\hat{y}_n, y_n) + \frac{1}{N} \sum_{n=1}^N l(\hat{y}_n \hat{z}_n, y_n z_n),
\end{equation}
where $l(\hat{y}_n, y_n) = - y_n \log (\hat{y}_n) - (1-y_n) \log (1 - \hat{y}_n)$.

\subsection{Embedding Masking}
\begin{figure}[!t]
\centering
\includegraphics[width=0.48\textwidth, trim = 0 0 0 0, clip]{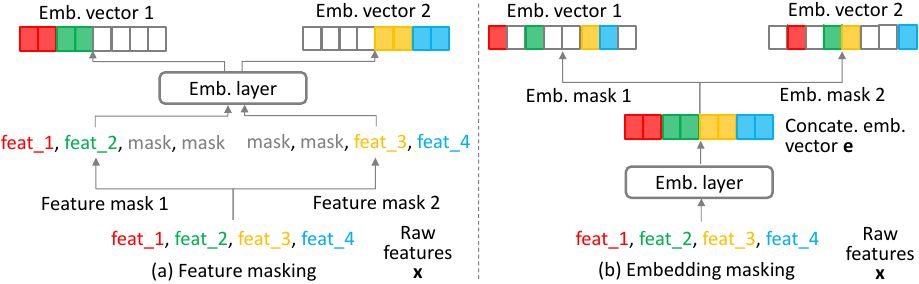}
\vskip -6pt
\caption{(a) Feature masking. (b) Embedding masking.}
\vskip -8pt
\label{fig_emb_mask}
\end{figure}

We now turn our attention to the CL task. Data augmentation is an important step that highly impacts the CL performance.
In \cite{yao2021self}, the authors propose to create two views of each original sample by feature masking for item recommendation (Fig. \ref{fig_emb_mask}(a)). The aim is that two differently masked views, each containing part of item features, can still well represent the same item.

However, feature masking does not work well in CVR prediction, because the input features are diverse, rather than only about the item. We cannot decide whether a user would like to convert on an ad if the ad features are masked. Therefore, we propose embedding masking (EM) in this paper, which is illustrated in Fig. \ref{fig_emb_mask}(b).

In EM, we apply two different element-wise masks on the concatenated long embedding vector $\mathbf{e}$ rather than on the raw features $\mathbf{x}$. Assume there are $F$ features and the embedding dimension for each feature is $K$. Then a feature mask has dimension $F$, but an embedding mask has dimension $FK$.
By EM, each masked view contains all (rather than part of) the features, except that some random embedding dimensions are masked. 
The aim is that the remaining embedding dimensions can still well represent the whole sample and the CL loss will force the learned embeddings to be more representative.
We denote the two augmented embedding vectors of the same sample as $\tilde{\mathbf{e}}_i$ and $\tilde{\mathbf{e}}_j$, which form a positive pair.

\subsection{Encoder and Traditional Contrastive Loss}
We map the two views $\tilde{\mathbf{e}}_i$ and $\tilde{\mathbf{e}}_j$ of the same sample to two high-level representation vectors $\mathbf{h}_i$ and $\mathbf{h}_j$ through the same encoder $f$.
That is, $\mathbf{h}_i = f(\tilde{\mathbf{e}}_i)$ and $\mathbf{h}_j = f(\tilde{\mathbf{e}}_j)$.
For simplicity, we use an MLP as the encoder, which contains several fully connected (FC) layers with the ReLU activation \cite{nair2010rectified} except the last layer (Fig. \ref{fig_model}(c)).

Given $N$ original samples in a mini-batch, there are $2N$ augmented samples. Given an anchor sample $\tilde{\mathbf{e}}_i$, the authors in \cite{chen2020simple} treat the other augmented sample $\tilde{\mathbf{e}}_j$ of the same original sample as the positive and treat other augmented samples as negatives. We illustrate it in Fig. \ref{fig_illus}(a).
The traditional contrastive loss \cite{chen2020simple} is %applied on the high-level representation vectors as
\begin{equation} \label{tra_cl_loss}
L_0 = - \frac{1}{2N} \sum_{i=1}^{2N} \log \frac{\exp\big(s(\mathbf{h}_i, \mathbf{h}_j)/\tau \big)}{\sum_{k \neq i} \exp\big(s(\mathbf{h}_i, \mathbf{h}_k)/\tau \big)},
\end{equation}
where $s(\mathbf{h}_i, \mathbf{h}_j)$ is the cosine similarity function and $\tau$ is a tunable temperature hyper-parameter.
This loss function aims to learn robust data representations such that similar samples are close to each other and random samples are pushed away in the latent space.

\subsection{False Negative Elimination}
\begin{figure}[!t]
\centering
\includegraphics[width=0.38\textwidth, trim = 0 0 0 0, clip]{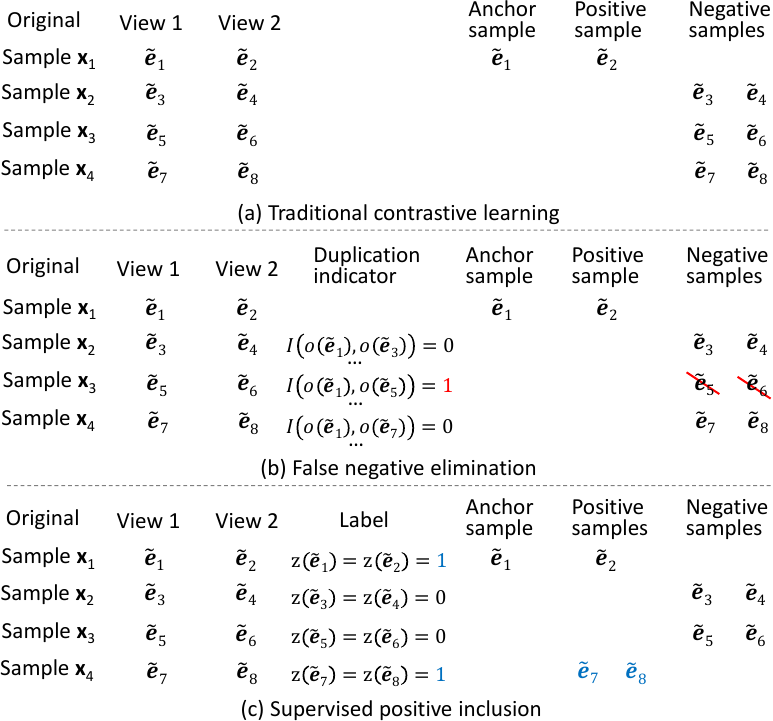}
\vskip -8pt
\caption{Anchor sample, positive sample(s) and negative samples in (a) traditional contrastive learning, (b) false negative elimination and (c) supervised positive inclusion.}
\vskip -10pt
\label{fig_illus}
\end{figure}

In advertising systems, it is common that an ad is shown to a user multiple times at different time epochs. Because user behaviors naturally contain uncertainty, it is possible that the user clicks the ad $a_1$ times and converts $a_2$ times ($a_2 \leq a_1$).
This results in $a_1$ click samples with the same features but possibly different conversion labels.
When such samples are included for CL, contradiction happens.
It is because augmented samples corresponding to original samples with different indices will be treated as negatives. However, their original samples actually have the same features.

Therefore, we propose a false negative elimination (FNE) component.
It generates a set $\mathcal{M}(i)$ for an anchor sample index $i$ (Fig. \ref{fig_illus}(b)). 
Note that, FNE only impacts the CL task and the supervised prediction model still uses all the original samples for training, as otherwise, the learned conversion probabilities are incorrect.
We use $o(\tilde{\mathbf{e}}_i)$ to denote the original sample of $\tilde{\mathbf{e}}_i$.
We introduce a duplication indicator where $I(o(\tilde{\mathbf{e}}_i), o(\tilde{\mathbf{e}}_k)) = 1$ indicates that $o(\tilde{\mathbf{e}}_i)$ and $o(\tilde{\mathbf{e}}_k)$ have the same features and it is 0 otherwise. Given an anchor sample index $i$, we define the set
\begin{equation}
\mathcal{M}(i) = \{j\} \cup \{k | I(o(\tilde{\mathbf{e}}_i), o(\tilde{\mathbf{e}}_k)) = 0 \}.
\end{equation}
$\mathcal{M}(i)$ contains the indices of samples that should be included in the denominator of the CL loss function.

\subsection{Supervised Positive Inclusion}
As the conversion label is sparse but also precious, we further propose a supervised positive inclusion (SPI) component to effectively leverage label information.
It generates a set $\mathcal{S}(i)$ with supervised positive included for an anchor sample index $i$.

Inspired by supervised contrastive learning \cite{khosla2020supervised}, we include additional positive samples for an anchor sample when its conversion label is 1 (Fig. \ref{fig_illus}(c)).
Note that in traditional contrastive learning \cite{chen2020simple}, an anchor sample $\tilde{\mathbf{e}}_i$ has a single positive sample $\tilde{\mathbf{e}}_j$.

Given an anchor sample index $i$, we define the set
\begin{equation}
\mathcal{S}(i) = \{j\} \cup \{k | z(\tilde{\mathbf{e}}_k) = z(\tilde{\mathbf{e}}_i) = 1, k \neq i, k \neq j\},
\end{equation}
where $z(\tilde{\mathbf{e}}_i)$ denotes the label of $\tilde{\mathbf{e}}_i$, which is the same as the original sample.
In other words, $\mathcal{S}(i) = \{j\}$ if $z(\tilde{\mathbf{e}}_i)=0$ (i.e., the anchor sample has label 0) and $\mathcal{S}(i)$ may contain more positive samples if $z(\tilde{\mathbf{e}}_i) = 1$.
We do not include supervised positive samples when $z(\tilde{\mathbf{e}}_i) = 0$ because of the data sparsity issue. It is possible that all the samples in a mini-batch has $z=0$, which makes all the samples supervised positives and there is no negative and no contrast at all.

\subsection{Contrastive Loss and Overall Loss}
$\mathcal{M}(i)$ generated by FNE and $\mathcal{S}(i)$ generated by SPI impact the contrastive loss. In particular, we define the contrastive loss used in this paper as
\begin{equation} \label{sup_cl_loss}
L_{cl} = - \frac{1}{2N} \sum_{i=1}^{2N} \left[ \frac{1}{|\mathcal{Q}(i)|} \sum_{q \in \mathcal{Q}(i)} \log \frac{\exp\big(s(\mathbf{h}_i, \mathbf{h}_q)/\tau \big)}{\sum_{k \in \mathcal{M}(i)} \exp\big(s(\mathbf{h}_i, \mathbf{h}_k)/\tau \big)} \right],
\end{equation}
where $\mathcal{Q}(i) = \mathcal{S}(i) \cap \mathcal{M}(i)$. For each anchor sample, we average over all its positives.
The overall loss is the combination of the supervised CVR prediction loss and the contrastive loss as
$L = L_{pred} + \alpha L_{cl}$, where $\alpha$ is a tunable balancing hyper-parameter.

\section{Experiments}

\subsection{Datasets}
\vskip -10pt
\begin{table}[!th]
\setlength{\tabcolsep}{2pt}
\renewcommand{\arraystretch}{1.05}
\caption{Statistics of experimental datasets.}
\vskip -10pt
\label{tab_stat}
\centering
\resizebox{0.48\textwidth}{!}{%
\begin{tabular}{|l|c|c|c|c|c|c|c|}
\hline
\textbf{Dataset} & \textbf{\# Fields} & \textbf{\# Train} & \textbf{\# Val} & \textbf{\# Test} & \textbf{\# Show} & \textbf{\# Click} & \textbf{\# Conv} \\
\hline
Industrial & 60 & 278.8M & 49.2M & 48.4M & 376.4M & 64.5M & 0.67M \\
\hline
Public & 17 & 2.3M & 0.98M & 3.3M & 6.6M & 3.3M & 0.018M \\
\hline
\end{tabular}}
\vskip -12pt
\end{table}

The statistics of the datasets are listed in Table \ref{tab_stat}. Both datasets contain samples from advertising systems with rich features, and are tagged with click and conversion labels. \textbf{1) Industrial dataset}: This dataset contains a random sample of user behavior logs from an industrial news feed advertising system in 2022.
\textbf{2) Public dataset}: This dataset is gathered from the traffic logs in Taobao\footnote{https://tianchi.aliyun.com/dataset/dataDetail?dataId=408}.

\subsection{Compared Methods}
We compare the following methods for CVR prediction. Base is the supervised prediction model. Other methods associate the same base model with different data regularization or CL algorithms.
\begin{itemize}
\item \textbf{Base}. The supervised CVR prediction model. In this paper, we use ESMM \cite{ma2018entire} as the base.
\item \textbf{FD}. Base model with random Feature Dropout \cite{volkovs2017dropoutnet} in the supervised task.
\item \textbf{SO}. Base model with Spread-Out regularization \cite{zhang2017learning} on original examples.
\item \textbf{RFM}. Random Feature Masking \cite{yao2021self}. Base model with a CL task. It randomly splits features into two disjoint sets.
\item \textbf{CFM}. Correlated Feature Masking \cite{yao2021self}. Base model with a CL task. It splits features according to feature correlation.
\item \textbf{CL4CVR}. The framework proposed in this paper.
\end{itemize}

\subsection{Settings}
\textbf{Parameter Settings.} We set the dimensions of fully connected layers in prediction towers and those in the CL encoder as \{512, 256, 128\}.
The training batch size is set to 64. All the methods are implemented in Tensorflow \cite{abadi2016tensorflow} and optimized by Adagrad \cite{duchi2011adaptive}. We run each method 3 times and report the average results.

\textbf{Evaluation Metric.}
The Area Under the ROC Curve (AUC) is a widely used metric for CVR prediction. It reflects the probability that a model ranks a randomly chosen positive sample higher than a randomly chosen negative sample. The larger the better.

\subsection{Experimental Results}

\subsubsection{\textbf{Effectiveness}}
\begin{table}[!t]
\setlength{\tabcolsep}{3pt}
\renewcommand{\arraystretch}{1.03}
\caption{Test AUCs on experimental datasets. The best result is in bold font. A \textnormal{small} improvement in AUC (e.g., 0.0020) can lead to a \textnormal{significant} increase in online CVR (e.g., 3\%). * indicates the statistical significance for $p \leq 0.01$ compared with the best baseline over paired t-test.}
\vskip -10pt
\label{tab_auc}
\centering
\begin{tabular}{|l|c c | c c|}
\hline
 & \multicolumn{2}{|c|}{\textbf{Industrial dataset}} & \multicolumn{2}{|c|}{\textbf{Public dataset}} \\
\hline
& CVR AUC & Gain & CVR AUC & Gain \\
\hline
Base & 0.8558 & - & 0.6524 & - \\
\hline
FD & 0.8452 & -0.0106 & 0.6469 & -0.0055 \\
\hline
SO & 0.8563 & +0.0005 & 0.6534 & +0.0010 \\
\hline
RFM & 0.8522 & -0.0036 & 0.6536 & +0.0012 \\
\hline
CFM & 0.8539 & -0.0019 & 0.6541 & +0.0017 \\
\hline
CL4CVR & \textbf{0.8637}$^*$ & \textbf{+0.0079} & \textbf{0.6590}$^*$ & \textbf{+0.0066} \\
\hline
\end{tabular}
\vskip -4pt
\end{table}

Table \ref{tab_auc} shows the AUCs of different methods. It is observed that FD performs worst because it operates on the supervised task. SO performs better than RFM and CFM on the industrial dataset, but they have comparable performance on the public dataset.
CFM performs better than RFM because it further considers feature correlation.
CL4CVR performs best on both datasets, showing its effectiveness to cope with the data sparsity issue and to improve the CVR prediction performance.

\subsubsection{\textbf{Ablation Study}}

\begin{table}[!t]
\setlength{\tabcolsep}{3pt}
\renewcommand{\arraystretch}{1.03}
\caption{Ablation study. Test AUCs on experimental datasets. EM - embedding masking. FNE - false negative elimination. SPI - supervised positive inclusion.}
\vskip -10pt
\label{tab_ablation}
\centering
\begin{tabular}{|l|c c | c c |}
\hline
 & \multicolumn{2}{|c|}{\textbf{Industrial dataset}} & \multicolumn{2}{|c|}{\textbf{Public dataset}} \\
\hline
& CVR AUC & Gain & CVR AUC & Gain \\
\hline
Base & 0.8558 & - & 0.6524 & - \\
\hline
EM & 0.8586 & +0.0028 & 0.6572 & +0.0048 \\
\hline
EM + FNE  & 0.8605 & +0.0047 & 0.6581 & +0.0057 \\
\hline
EM + SPI & 0.8617 & +0.0059 & 0.6580 & +0.0056 \\
\hline
EM + FNE + SPI & \textbf{0.8637} & \textbf{+0.0079} & \textbf{0.6590} & \textbf{+0.0066} \\
\hline
\end{tabular}
\vskip -6pt
\end{table}

Table \ref{tab_ablation} lists the AUCs of three components in CL4CVR.
It is observed that EM itself outperforms RFM and CFM, showing that embedding masking is more suitable than feature masking for CVR prediction. The incorporation of the FNE component or the SPI component leads to further improvement.
CL4CVR that uses all the three components perform best, showing that these components complement each other and improve the prediction performance from different perspectives.

\subsubsection{\textbf{Impact of the Temperature and the CL Loss Weight}}
\begin{figure}[!t]
\centering
\subfigure[Industrial dataset]{\includegraphics[width=0.23\textwidth, trim = 0 0 0 0, clip]{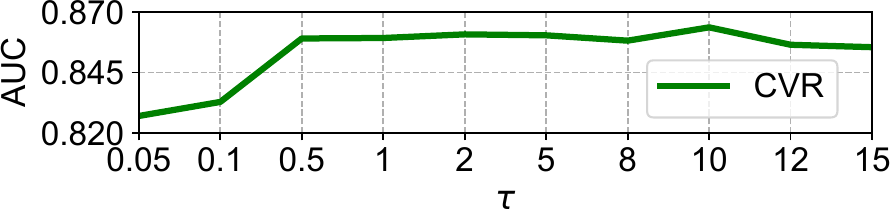}}
\subfigure[Public dataset]{\includegraphics[width=0.23\textwidth, trim = 0 0 0 0, clip]{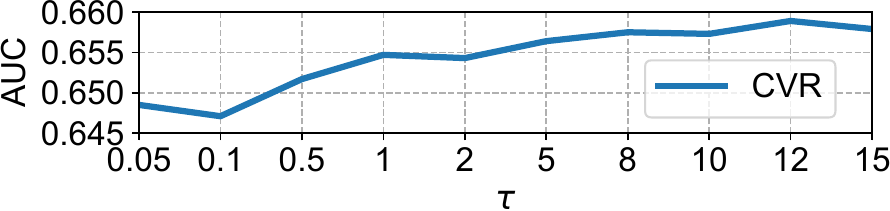}}
\vskip -16pt
\caption{Impact of the temperature $\tau$.}
\vskip -14pt
\label{fig_tau}
\end{figure}

\begin{figure}[!t]
\centering
\subfigure[Industrial dataset]{\includegraphics[width=0.23\textwidth, trim = 0 0 0 0, clip]{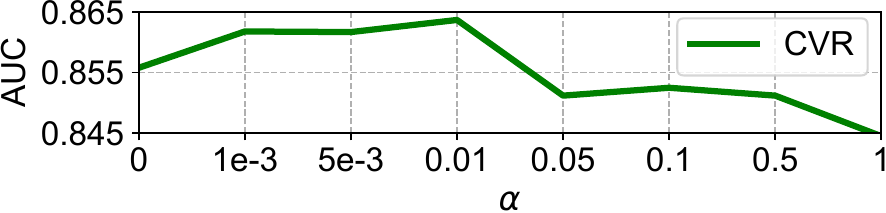}}
\subfigure[Public dataset]{\includegraphics[width=0.23\textwidth, trim = 0 0 0 0, clip]{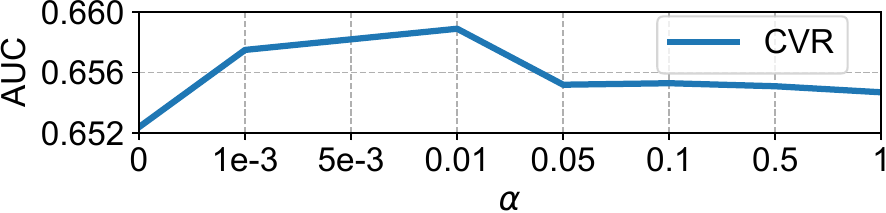}}
\vskip -16pt
\caption{Impact of the CL loss weight $\alpha$.}
\vskip -16pt
\label{fig_alpha}
\end{figure}

Fig. \ref{fig_tau} plots the impact of the temperature $\tau$. It is observed that generally a large $\tau$ works well on the two datasets. Fig. \ref{fig_alpha}
plots the impact of the CL loss weight $\alpha$, where 0 denotes the supervised base model. It is observed that when $\alpha$ increases initially, performance improvement is observed. But when $\alpha$ is too large, too much emphasis on the CL task will degrade the performance.

\section{Related Work}
\textbf{CVR prediction.}
The task of CVR prediction \cite{lee2012estimating,chapelle2014modeling,lu2017practical} in online advertising is to estimate the probability of a user makes a conversion event on a specific ad.
\cite{lee2012estimating} estimates CVR based on past performance observations along data hierarchies. \cite{chapelle2014simple} proposes an LR model and \cite{agarwal2010estimating} proposes a log-linear model for CVR prediction.
\cite{rosales2012post} proposes a model in non-guaranteed delivery advertising.
\cite{ma2018entire} proposes ESMM to exploit click and conversion data in the entire sample space.
ESM$^2$ \cite{wen2020entire}, GMCM \cite{bao2020gmcm} and HM$^3$ \cite{wen2021hierarchically} exploit additional purchase-related behaviors after click (e.g., favorite, add to cart and read reviews) for CVR prediction.

\textbf{Contrastive learning.}
Contrastive learning \cite{chen2020simple,yu2022self,gao2021simcse} offers a new way to conquer the data sparsity issue via unlabeled data. It is able to learn more discriminative and generalizable representations.
Contrastive learning has been applied to a wide range of domains such as computer vision \cite{chen2020simple}, natural language processing \cite{gao2021simcse} and recommendation \cite{yu2022self}. In the recommendation domain, most data augmentation methods are ID-based sequence or graph approaches \cite{xie2022contrastive,wu2021self,yu2022self,xia2022hypergraph}, which do not apply to CVR prediction which is a feature-rich problem. The most relevant work is \cite{yao2021self} which proposes feature masking for item recommendation.

\section{Conclusion}
In this paper, we propose the Contrastive Learning for CVR prediction (CL4CVR) framework. It associates the supervised CVR prediction task with a contrastive learning task, which can learn better data representations exploiting abundant unlabeled data and improve the CVR prediction performance. To tailor the contrastive learning task to the CVR prediction problem, we propose embedding masking, false negative elimination and supervised positive inclusion strategies. Experimental results on two real-world conversion datasets demonstrate the superior performance of CL4CVR.

\bibliographystyle{ACM-Reference-Format}
\balance
\bibliography{ref}

\end{document}